\PassOptionsToPackage{table,xcdraw}{xcolor}
\documentclass[sigconf]{acmart}
\usepackage{courier}
\usepackage[ruled,vlined]{algorithm2e}
\usepackage{graphicx}
\usepackage{paralist}
\usepackage{tabularx}
\usepackage{balance}
\usepackage{multirow}
\usepackage{hyperref}
\usepackage{multicol}
\usepackage{pbox}
\usepackage{mathtools}
\usepackage[TABBOTCAP]{subfigure}
\usepackage{algorithmic}
\usepackage{paralist}
\usepackage{tabularx}
\usepackage{url}
\usepackage{balance}
\usepackage{multirow}
\usepackage{multicol}
\usepackage{amsmath}
\usepackage{setspace}
\usepackage{xcolor}
\usepackage{enumitem}
\usepackage{tcolorbox}
\usepackage{verbatim}
\usepackage{float}
\usepackage[normalem]{ulem}
\usepackage{xspace}
\usepackage{ifthen}
\usepackage{pifont}
\usepackage{tikz}
\usepackage{adjustbox}
\usepackage{cleveref} 

\input{macros}

\AtBeginDocument{%
  \providecommand\BibTeX{{%
    \normalfont B\kern-0.5em{\scshape i\kern-0.25em b}\kern-0.8em\TeX}}}

\copyrightyear{2024}
\acmYear{2024}
\setcopyright{rightsretained}
\acmConference[ICSE '24]{2024 IEEE/ACM 46th International Conference on Software Engineering}{April 14--20, 2024}{Lisbon, Portugal}
\acmBooktitle{2024 IEEE/ACM 46th International Conference on Software Engineering (ICSE '24), April 14--20, 2024, Lisbon, Portugal}\acmDOI{10.1145/3597503.3639163}
\acmISBN{979-8-4007-0217-4/24/04}

\definecolor{main}{HTML}{5989cf}    
\definecolor{sub}{HTML}{cde4ff}     
\newtcolorbox{BugReportBox}[1][!ht]{
    colback = sub, 
    colframe = main, 
    boxrule = 0pt, 
    toprule = 4pt 
}

\begin{document}

\title{Semantic GUI Scene Learning and Video Alignment for Detecting\\ Duplicate Video-based Bug Reports}

\author{Yanfu Yan}
\email{yyan09@wm.edu}
\affiliation{
  \institution{William \& Mary}
  \city{Williamsburg}
  \state{Virginia}
  \country{USA}
}

\author{Nathan Cooper}
\email{nacooper01@.wm.edu}
\affiliation{
  \institution{William \& Mary}
  \city{Williamsburg}
  \state{Virginia}
  \country{USA}
}

\author{Oscar Chaparro}
\email{oscarch@wm.edu}
\affiliation{
  \institution{William \& Mary}
  \city{Williamsburg}
  \state{Virginia}
  \country{USA}
}

\author{Kevin Moran}
\email{kpmoran@ucf.edu}
\affiliation{
  \institution{University of Central Florida}
  \city{Orlando}
  \state{Florida}
  \country{USA}
}

\author{Denys Poshyvanyk}
\email{denys@cs.wm.edu}
\affiliation{
  \institution{William \& Mary}
  \city{Williamsburg}
  \state{Virginia}
  \country{USA}
}

\renewcommand{\shortauthors}{Yan, et al.}

\begin{abstract}
Video-based bug reports are increasingly being used to document bugs for programs centered around a graphical user interface (GUI). 
However, developing automated techniques to manage video-based reports is challenging as it requires identifying and understanding often nuanced visual patterns that capture key information about a reported bug.  
In this paper, we aim to overcome these challenges by advancing the bug report management task of \textit{duplicate detection} for video-based reports. To this end, we introduce a new approach, called \approach, that adapts the scene-learning capabilities of vision transformers to capture subtle visual and textual patterns that manifest on app UI screens --- which is key to differentiating between similar screens for accurate duplicate report detection. \approach also makes use of a video alignment technique capable of adaptive weighting of video frames to account for typical bug manifestation patterns.  
In a comprehensive evaluation on a benchmark containing 7,290 duplicate detection tasks derived from 270 video-based bug reports from 90 Android app bugs, the best configuration of our approach achieves an overall mRR/mAP of 89.8\%/84.7\%, and for the large majority of duplicate detection tasks, outperforms prior work by $\approx$9\% to a statistically significant degree. 
Finally, we qualitatively illustrate how the scene-learning capabilities provided by \approach benefits its performance.
\end{abstract}



\maketitle

\section{Introduction}
\label{sec:intro}

Video-based bug reports are becoming increasingly popular for mobile applications~\cite{Feng:ICSE'22,kuramoto2022visual,feng2023read,chen2022extracting}. As mobile app bugs typically manifest visually via the graphical user interface~(GUI), recording videos depicting bugs is more natural compared to textual bug reports~\cite{feng2023read,chen2022extracting}. App users can easily record app bugs via the recording features of mobile operating systems (\eg Android~\cite{android_video_feature}) or via third-party recording apps~\cite{android_screen_recording}. Additionally, popular issue trackers, such as GitHub~\cite{github_videos}, offer easy-to-use features for users to submit these videos to app developers. The rapidly increasing use of videos in mobile app issue trackers has been documented by recent studies~\cite{Feng:ICSE'22,kuramoto2022visual}. Feng \etal studied open source apps hosted on FDroid~\cite{fdroid}, and reported the usage of over 13k video recordings in issue trackers between 2012-2020, with a significant increase in usage during 2018-2020 (\ie a 15\% - 35\% increase). Kuramoto~\etal~\cite{kuramoto2022visual} reported a 13\% increase in issues containing videos in 2017-2021 for 289k popular GitHub projects.

While video-based bug reporting offers various advantages (ease of recording and submission, and visual details about app bugs~\cite{Feng:ICSE'22,kuramoto2022visual,feng2023read,chen2022extracting,cooper2021takes}), it also presents several challenges for developers during bug report management tasks, particularly in scenarios where a high volume of bug reports is encountered~\cite{Feng:ICSE'22,kuramoto2022visual,feng2023read,chen2022extracting}.

One of the most challenging tasks for developers is determining whether video-based bug reports depict the same app bug. This situation arises when multiple users independently report identical problems with the application (\eg during crowd-sourced app testing~\cite{hao2019ctras,du2022semcluster,cooper2021takes}). In such scenarios, developers face the challenge of watching, understanding, and assessing incoming and previously submitted video-based bug reports. This task can be extremely challenging since these recordings typically show numerous steps executed rapidly, making it difficult to recognize the bug reproduction scenario from the videos~\cite{chen2022extracting,feng2023read,cooper2021takes}. Additionally, the depicted buggy app behavior may not be apparent in the videos for the various types of bugs that apps can show in their GUI~\cite{escobar2019mutapk}. Developers often need to pause and replay the videos multiple times in order to fully understand the reported problems~\cite{chen2022extracting,feng2023read}. The task of duplicate (video-based) bug report detection is crucial during the bug triage process, as it helps developers avoid excessive redundant effort in investigating and resolving identical issues~\cite{Wang2019,hao2019ctras,du2022semcluster,cooper2021takes}.

This challenge is particularly prominent in crowd-sourced testing of mobile apps~\cite{hao2019ctras,du2022semcluster}, wherein software vendors engage a large distributed user base to test applications across diverse operational environments, for example, encompassing various devices, locations, and mobile networks. Crowd-sourced app testing often leads to multiple users encountering and reporting the same app-related issues.  In fact, previous research has found that a substantial proportion (80\%+) of bug reports submitted by users during crowd-sourced app testing are duplicates~\cite{Wang2019}. Consequently, developers often spend considerable effort on duplicate detection, which can impede the overall bug resolution process~\cite{Wang2019,hao2019ctras,du2022semcluster,cooper2021takes}. 

In this paper, we propose \approach, a novel automated approach designed to assist developers in identifying duplicate video-based bug reports. \approach combines visual representation learning, information retrieval, and sequence-based algorithms, to analyze the visual, textual, and sequential information present in video-based bug reports. Through these methods, \ap establishes the degree of similarity between videos in reporting the same bug, thus enabling the automated detection of duplicate reports.

To model the visual information within videos, \approach leverages the Vision Transformer (ViT) architecture~\cite{dosovitskiy2020image} and the self-supervised training scheme DINO~\cite{caron2021emerging}, which extract rich hierarchical features that explicitly capture scene layout information related to GUI screens. In addition, \ap analyzes the textual content of videos by leveraging the Efficient and Accurate Scene Text Detector (EAST)~\cite{zhou2017east} and a Transformer-based Optical Character Recognition (TrOCR) model~\cite{li2021trocr}, which accurately localize and extract text from video frames. By encoding this textual content via an adapted vector space model (VSM)~\cite{Gospodnetic2004LuceneIA}, \approach assesses the textual similarity between two videos. Finally, to encode the sequential aspect of videos, \approach incorporates an adapted version of the classical longest common substring algorithm, giving higher weight to subsequent video frames that show the buggy app behaviors even if the videos show distinct bug reproduction scenarios.

We evaluate \approach using a comprehensive benchmark of 7,290 duplicate detection tasks, constructed from 270 video-based bug reports representing 90 unique bugs found in nine Android apps. We created this benchmark by extending an existing dataset that relied mostly on synthetic bugs~\cite{cooper2021takes}. Specifically, we extended it by incorporating 90 video-based bug reports pertaining to 30 real bugs of different kinds (\eg crashes, incorrect app output, and cosmetic issues) from three additional apps, resulting in a more comprehensive, realistic, and diverse benchmark.

Through multiple ablation experiments, we systematically assess the performance of the individual components of \approach as well as various combinations of these components. Our evaluation demonstrates that the most optimal configuration of \approach (when visual, textual, and sequential video information is combined) achieves an overall mRR/mAP of 89.8\%/84.7\%, surpassing the performance of an existing duplicate detector by $\approx$9\% (with statistical significance). These results suggest
that \ap can significantly reduce the effort required to identify duplicate video-based bug reports, as developers would only need to review fewer video reports to assess whether an incoming report depicts a known bug.

Furthermore, we conducted a qualitative analysis to understand the reasons behind \approach' performance compared to prior work. Notably, \approach exhibits an interpretable representation of video frames, effectively capturing nuanced patterns related to GUI component style, composition, and layout, which are crucial in accurately distinguishing duplicate video-based bug reports. 

In summary, this paper makes the following main contributions:
\begin{itemize}[leftmargin=2.0em] 
    \item A new approach (\ap) that leverages visual representation learning for the graphical/lexical analysis of video-based bug reports. It also leverages sequential information within these reports for more accurate duplicate detection; 
    \item A comprehensive empirical evaluation that demonstrates state-of-the-art improvements achieved by \approach in detecting duplicate video-based bug reports, when compared to a prior duplicate detector;
    \item A qualitative analysis into potential reasons for our improved results as well as future research directions on how the results can be further improved; and
    \item A publicly available benchmark for evaluating duplicate video-based bug report detectors~\cite{janus-tool}, composed of 7,290 duplicate detection tasks created from 120/150 reports about 40/50 real/synthetic bugs---the largest benchmark to date.
    \looseness=-1
\end{itemize}
\section{The J{\large ANUS} Duplicate Detector}
\label{sec:approach}

This section describes the architecture and design details behind \ap, our approach for duplicate video-based bug report detection.

\subsection{Problem Formulation and Challenges}
\label{sec:problem}

We formulate the problem of duplicate detection as an information retrieval (IR) problem, as is typically done for textual bug reports~\cite{zhang2023duplicate,Lin2016a,Kang2017}. A newly-submitted video-based bug report (the query) is compared against the set of previously submitted video reports in the issue tracker (the corpus) via a retrieval engine (\eg~\ap), which retrieves and ranks the corpus reports by their similarity with the query. The higher a video-based report is ranked, the more likely it is to depict the same bug as the query. A developer then would watch the ranked videos in a top-down
manner, marking the new video as duplicate if  they find a video
depicting the same bug. Video-based bug reports depict the incorrect behavior of an application (\eg GUI screens showing a crash, layout problems, or functional misbehavior), and the actions performed by the user on the GUI screens that lead to such misbehavior (\ie the steps to reproduce the bug). Duplicate video-based bug reports are pairs of two reports, \eg the query report and one corpus report, that depict the same buggy behavior, possibly showing different GUI steps, as multiple sequences of steps can lead to the manifestation of the same buggy behavior.
An advantage of an IR  formulation over other methods (\eg binary classification as (non-)duplicates  reports~\cite{chaparro2019reformulating}) is the fact that a ranked list  
gives higher flexibility to developers, because multiple bug reports are recommended as possibly showing the same bug.

While the primary goal of a duplicate detector is to identify whether two distinct videos depict the same incorrect app behavior, there are multiple challenges that make this task particularly difficult. For instance,  duplicate videos may vary in length and display different reproduction steps, stemming from diverse reproduction scenarios executed by the users or the omission of certain steps during recording. Even if the reproduction steps appear the same or highly similar across videos, users may execute them at varying speeds. Distinguishing between different videos displaying distinct yet similar unexpected app behavior and reproduction steps can pose challenges for detectors. Furthermore, certain applications may exhibit dynamic content. For example, a mobile web browser allows users to navigate websites with varying layouts/content.

\begin{figure}[t] 
\centering 
\includegraphics[width=0.48\textwidth]{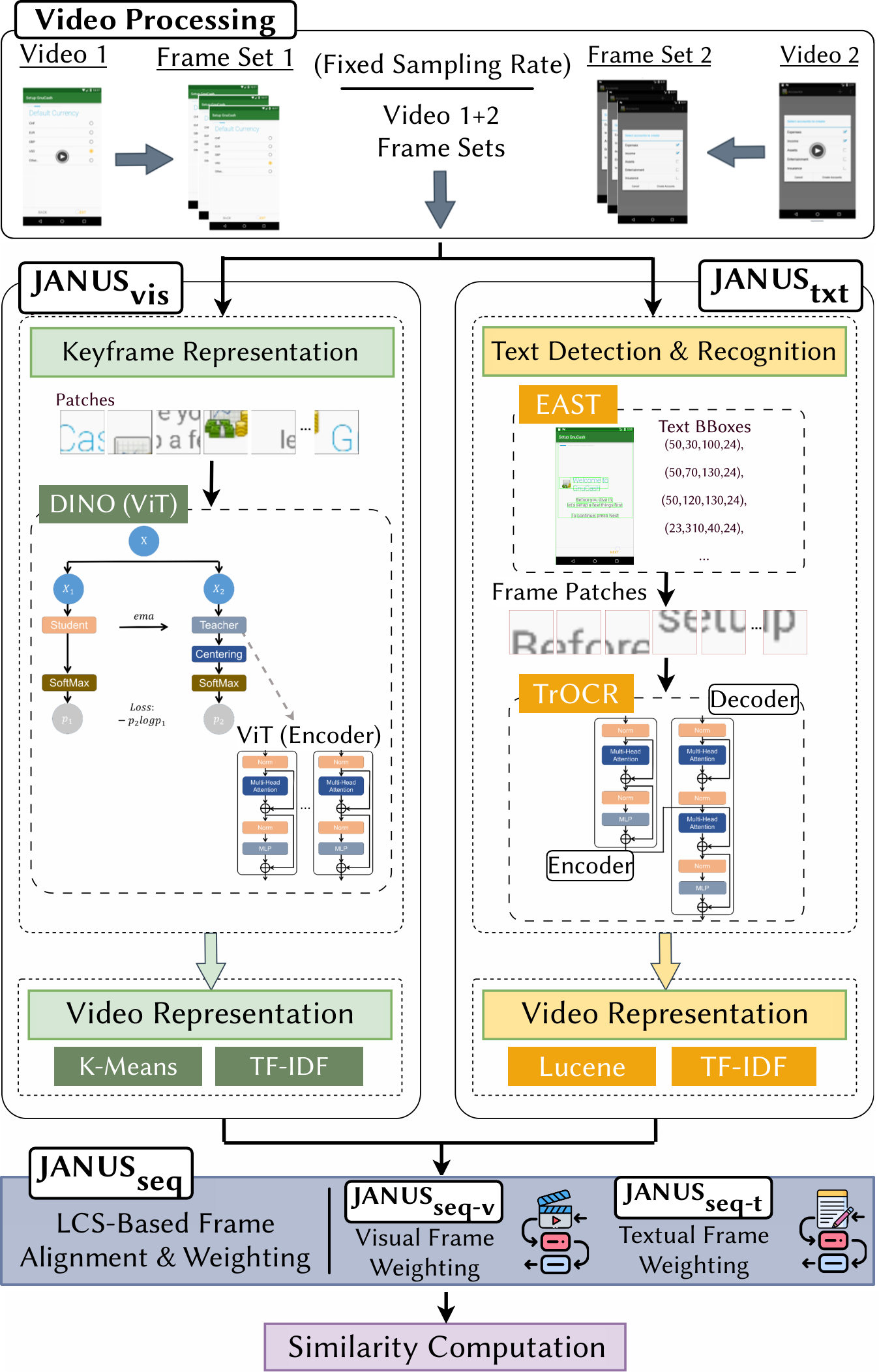}
\vspace{-2.0em}
\caption{\normalsize{Overview of the \approach duplicate detector.}}
\label{fig:approach}
\end{figure}

\vspace{-1em}
\subsection{\ap Overview}
\label{sec:approach_overview}

An overview of \toolname is shown in Fig. \ref{fig:approach}. \ap receives as input two video-based bug reports and outputs a similarity score that indicates how similar they are at depicting the same app bug. 
\ap can be used to compute scores between a new video-based bug report and a corpus of videos representing previously-submitted bug reports. The scores allow for ranking the corpus videos as a list of potential duplicate candidates. The goal of \ap is to rank higher in this list, the actual duplicates for the new video. 

Internally, \toolname begins sampling a number of frames from the two videos at a given rate (every sixth frame following the findings of past work~\cite{cooper2021takes}) to reduce overhead, given that successive frames tend to be exact or near duplicates of each other. Next, \ap computes a vector representation of the videos, by processing the visual and/or textual content of the frames.  \aps visual component, \janusv, vectorizes each video into a visual TF-IDF representation by discretizing the frames into a Bag of Visual Words (BoVW)~\cite{jiang2007towards}, using a feature extractor based on a Vision Transformer (ViT) model~\cite{dosovitskiy2020image} and the DINO self-supervised training scheme~\cite{caron2021emerging}. \aps textual component, \janust, vectorizes each video into a textual TF-IDF representation by extracting video frame text (via the EAST~\cite{zhou2017east} and TrOCR~\cite{li2021trocr} models) and constructing a document of the concatenated text, represented as a Bag of Words (BoW)~\cite{Salton:TFIDF86}. Each pair of visual or textual TF-IDF representations is then compared via cosine similarity. The visual and textual similarities can be used individually to rank duplicate candidates, or they can be combined into a single similarity score to account for both modalities of information, ideally leading to more effective duplicate detection. 

To account for the sequential nature of video-based bug reports, which typically show the reproduction steps first and the incorrect app behavior afterward, \ap can compute an alternative similarity score, based on a customized version of the longest common substring (LCS) algorithm, which matches the vector representation of video frames via cosine similarity and produces an overall similarity score that weights more heavily the later frames in the video than the earlier ones. This similarity is computed by \aps sequential component, \apss, which operates on the visual (\apsv) and textual (\apst) vector representations of the frames.

\subsection{\apv: Visual Representation of Videos}
\label{sec:visual_approach}

\apv obtains a visual representation of a video in two steps. First, the sampled video frames are resized to $224\times224$ (\textit{pixels}) and encoded via visual representation learning~\cite{kolesnikov2020big}. Second, these frame embeddings are further processed into a Bag of Visual Words (BoVW)~\cite{kordopatis2019fivr}, which is used to represent a video as a TF-IDF vector~\cite{Salton:TFIDF86}. The goal is to learn useful visual information from app GUI components and layouts shown in the videos to distinguish potential duplicates from non-duplicates.

\subsubsection{\textbf{Visual Representation of Video Frames}}

Visual representation learning aims to obtain high-quality visual representations that are helpful for downstream tasks such as image classification~\cite{Krizhevsky:NIPS12}, object detection~\cite{zhao2019object}, or image captioning~\cite{hossain2019comprehensive}. This task is typically carried out in an unsupervised, self-supervised, or supervised manner~\cite{chen2020simple, radford2021learning}. Most recently, there has been a focus on contrastive~\cite{chen2020simple, radford2021learning} and distillation learning methods~\cite{caron2021emerging}. A promising technique, known as the Vision Transformer (ViT)~\cite{dosovitskiy2020image} has recently been proposed to better learn visual representations. The performance of this architecture has been demonstrated to surpass or, at the very least, match previous models relying on Convolutional Neural Networks (CNNs) for image classification. However, the most significant advantage of ViT lies in its ability to excel beyond CNNs in capturing explicit information concerning the semantic segmentation of an image (\ie layouts and object boundaries)~\cite{dosovitskiy2020image}.

We posit that learning object segmentation within an image is particularly useful for app GUI screens, given their structured, component-based nature. Hence, we adopted the ViT architecture for designing \apv. 
The ViT architecture is comprised of a standard Transformer encoder model~\cite{devlin2018bert} but instead of lexical tokens, ``patches'' from images are fed into the network. These patches are treated the same way that tokens are in lexical transformers: they are linearly transformed and have added positional embeddings. 

Given that image-level supervision requires labor-intensive annotations and limits the information that can be learned during training to a single concept with a few categories of objects (as is the case of app GUI screens, which contain components and layouts of well-defined kinds), we need to train our ViT model in a self-supervised manner. \apv trains its ViT using the self-supervised training methodology DINO~\cite{caron2021emerging}, which leverages a student-teacher knowledge distillation training scheme ~\cite{Hinton2015DistillingTK}.
In this scheme, 
the student network is trained to match the distribution of the teacher network by minimizing the standard cross-entropy loss. Usually, the teacher network is larger than the student network in terms of the number of model parameters. However, the teacher network in DINO is built from the past iterations of the student network with an exponential moving average strategy, whose parameters are frozen over an epoch by applying a stop-gradient operator, given that direct replication of the student weights fails to converge. The outputs of both networks are normalized using a temperature softmax. To adapt the knowledge distillation architecture to self-supervised learning, two global views and several local views are constructed on the basis of data augmentations \cite{Grill2020BYO} and the multi-crop strategy \cite{Caron2020Multicrop}, with local views passed through the student while only the global views are passed through the teacher network, to encourage local-to-global correspondence. By combining DINO with ViT, we aim to further improve the ability to capture global GUI layouts.

Through this self-supervised training process, the model learns a rich representation of images that emphasize scene layouts and object boundaries. To further refine the DINO model's capabilities to our domain of app GUI screens, we fine-tuned  \aps ViT model, which was pre-trained on ImageNet~\cite{deng2009imagenet}, on a collection of 66k mobile app screenshots from the popular RICO dataset \cite{deka2017rico}. We directly use the projected output of the \emph{[CLS]} token, a special token that marks the aggregation of all image patch embeddings, from the last block of the ViT model as the representation of video frames. 

\subsubsection{\textbf{Visual Representation of Videos}}

To represent a video, \apv implements a BoVW + TF-IDF approach since it has been shown to be more useful for video retrieval compared to other approaches~\cite{kordopatis2019fivr} (\eg using directly the frame representations for similarity computation or aggregating them into a single vector).

\ap discretizes the frame representations by leveraging a Codebook of visual words~\cite{kordopatis2019fivr}. The Codebook represents a catalog of visual words, which are representative vectors found in a corpus of images (in our case, images of app GUIs). The Codebook is constructed via a trained $K$-Means model that clusters the corpus of image representations into $K$ clusters, the centroids being the visual words. \ap then assigns each video frame representation to its closest cluster centroid (\ie a visual word) via Euclidean distance. The Codebook is trained by randomly sampling 15k mobile app screenshots from the RICO dataset~\cite{deka2017rico}, vectorizing them via our fine-tuned ViT model, and running the $K$-Means algorithm on the vectors, with $K=1k$ recommended by prior work~\cite{kordopatis2019fivr}. We take a sample rather than using the full RICO dataset due to computational constraints of the $K$-Means algorithm. The Codebook is trained only once before the TF-IDF representation approach is applied.

Once each frame representation is discretized to its corresponding visual word, \ap computes a TF-IDF vector representation of a video, as similarly done for text retrieval~\cite{Salton:TFIDF86}. The term frequency (TF) is the count of each visual word in the video. The inverse document frequency (IDF) is the count of BoVW representations of existing videos where a visual word appears. Since a corpus of existing videos for a particular app may be small and may lack diversity, we consider the set of RICO images as the corpus of existing videos. By considering the diversity of apps in the RICO dataset, we aim to improve the generalization of the TF-IDF video representations.

\apv compares the TF-IDF representation of two videos via cosine similarity to establish the likelihood of the videos showing the same app bug.  This method is applied to the existing corpus of TF-IDF visual representations for an app to generate a ranked list of candidate duplicate videos for a new video-based bug report. 

To address potential biases due to random sampling when creating the Codebook, we adapted \apv to use four Codebooks (each trained on 15k RICO images, 60k in total). Specifically, \apv uses each Codebook to produce similarity scores for a set of videos. These similarity scores are averaged to produce a final set of similarities and video ranking. More details are given in \cref{sec: experiment_config}.

\subsection{\janust: Textual Representation of Videos}
\label{sec:textual_approach}

\apt creates a textual representation of a video in two  steps: (1) it localizes and extracts the text present in video frames via neural text localization and Optical Character Recognition (OCR); and (2), it encodes the extracted text using a standard TF-IDF representation~\cite{Salton:TFIDF86}. The goal is to leverage the text from labels, messages, and other sources shown in the frames to compute video similarity.

For the first step, \apt has two components: (1) a text localization component that proposes image regions where text is rendered, and (2) a text recognition component that takes those regions and extracts any text present in them. The text localization component implements the Efficient and Accurate Scene Text Detector (EAST) model~\cite{zhou2017east}, which has been trained to directly derive region proposals. The text recognition component leverages the TrOCR Transformer model~\cite{li2021trocr}, which takes region proposals from EAST and directly predicts the text represented in the proposals. The combination of EAST and TrOCR was adopted over the popular TesseractOCR~\cite{tesseract-ocr} approach because: (1) such a combination simplifies the overall OCR pipeline since it relies on neural models only, without needing heuristic-based approaches to filter out poor text region candidates (as TesseractOCR does); and (2) such a combination has shown strong performance improvements in detecting scene text as well as handwritten/printed text, which means it is less sensitive to noise in the images. Each video frame is put through this 2-stage pipeline to extract its text. 

For the second step, \apt concatenates the text from all video frames and pre-processes it via tokenization, lemmatization, and removal of special characters, such as non-ASCII characters, punctuation, or stop words.  This resulting text is used to build a Bag of Words (BoW) representation of the video, which is then encoded as a standard textual TF-IDF representation using the popular Lucene library~\cite{Gospodnetic2004LuceneIA}, which implements the standard information retrieval Boolean model and the Vector Space Model (VSM)~\cite{Salton:TFIDF86}. We use this textual representation approach over neural text encoding models because it is based on exact text matching, which could lead to more accurate similarity computation of duplicate videos (as they are likely to show the same text on the buggy app screens).

Finally, \apt compares the TF-IDF representation of two videos using Lucene's similarity scoring function (based on cosine similarity and document length normalization)~\cite{lucene-tfidfsimilarity}. Similarity computation can be applied to a corpus of video-based bug reports to generate a ranked list of potential duplicate videos to the new video.

\subsection{\apss: Sequential Similarity of Videos}
\label{sec:sequential_approach}

\apv and \apt ignore the sequential order of the videos, as these components are based on Bags Of (Visual) Words. However, the buggy app behavior is typically shown toward the end of a video-based bug report, after the bug reproduction steps have been rendered. To account for the sequential order of the videos, \ap employs a modified version of the longest common substring (LCS) algorithm to compute an alternative similarity score between videos. This approach is coined as \apss and operates on both visual (\apsv) and textual representations of the videos (\apst). 

\apss treats a video as a sequence of visual/textual words, based on the vector representation of the video frames, and applies an LCS-based approach for similarity computation. Intuitively, the longer the LCS between videos is, the higher their similarity is. The textual representation of a video frame is the TF-IDF vector of the text extracted from the frame, using the approach described in \cref{sec:textual_approach}. In the standard word-based LCS algorithm, words are compared using exact text matching. To account for similar, yet different video words (which might be common for textual video representations), we relaxed this matching scheme and instead used cosine similarity between video frame representations. Additionally, the similarity-based matching should weigh more heavily the frames that appear later in the videos as they are more likely to show the buggy app behavior and should give a normalized similarity score between zero and one.

Given these requirements, we defined the following similarity computation for \apss: $S_{seq} = \frac{\textrm{w-}LCS}{\textrm{max w-}LCS}$, 
where the numerator, w-$LCS$, represents the amount of overlap between two videos, given by our modified LCS algorithm, which uses the cosine similarity between frames (rather than exact matching) and a weighting scheme that favors later frames in the videos. The weighting scheme is  $\frac{i}{m} \times \frac{j}{n}$, where $i$ is the $ith$ frame of a first video, with $m$ being its \# of frames, and  $j$ is the $jth$ frame of a second video, with $n$ being its \# of frames. The denominator, max w-$LCS$, represents the maximum possible overlap if the videos were identical. Since the videos could be of different lengths, we align the end of the shorter video (with length $min$), to the end of the longer video (with length $max$), and calculate the maximum overlap as: $\sum_{i=1}^{min}\frac{i}{min}\times\frac{max - i}{max}$.

\subsection{Combining \aps Components}
\label{sec:approach_combinations}

To design \ap, we explore different combinations of its components. The similarity scores from \apv and \apt can be linearly combined as $(1 - w) \times S_{vis} + w \times S_{txt}$, with $w \in [0, 1]$--- the higher $w$ is the more weight it gives to textual information from the videos. We also explore various combinations that replace this similarity calculation with those given by \apss (\apsv \& \apst), which consider the sequential video information.

\section{Evaluation Methodology}
\label{sec:evaluation}

We investigate the performance of \approach's components (\janust, \janusv, and \apss), as well as the performance of various combinations of these components, and compare these to a baseline duplicate detection technique proposed in prior work~\cite{cooper2021takes}. Additionally, we aim to understand why we observe various trends in the overall performance of \approach, and qualitatively examine cases where \ap is able to outperform the baseline technique.

To that end, we formulate the following research questions (RQs):

\begin{enumerate}[label=\textbf{RQ$_\arabic*$:}, ref=\textbf{RQ$_\arabic*$}, wide, labelindent=0pt]\setlength{\itemsep}{0.2em}
    \item \label{rq:visual}{\textit{What is \apvs duplicate detection performance?}}
    \item \label{rq:textual}{\textit{What is \apts duplicate detection performance?}}
    \item \label{rq:sequential}{\textit{What is \apsss duplicate detection performance?}}
    \item \label{rq:combined}{\textit{What is the performance of \ap's component combinations?}}
\end{enumerate}

\subsection{Duplicate Detection Dataset}

We constructed a comprehensive evaluation dataset by extending a prior dataset that mostly relied on synthetic app bugs~\cite{cooper2021takes}. The previous dataset collected 60 distinct bugs (35 crashes and 25 non-crashes) across six Android apps of different sizes and domains (\eg podcast, finance, and weight management apps).
The dataset contains ten confirmed real bugs 
and 50 bugs injected by the mutation testing tool MutAPK \cite{escobar2019mutapk}, which generates code mutations based on diverse mutant operators that affect various app features. The dataset includes three duplicate videos per bug, for a total of 180 video-based bug reports, and a set of 810 duplicate detection tasks per app, for a total of 4,860 tasks, created from the videos. We refer to this dataset as the \textit{original dataset}. We next describe how we extended this dataset and detail the creation of video-based bug reports and duplicate detection tasks to evaluate \ap.

\subsubsection{\textbf{Extended Real Bug Dataset}}

We extended the prior dataset by constructing an evaluation dataset containing \textit{only} real bugs. Wendland \etal \cite{wendland2021andror2} released the AndroR2 dataset containing 90 manually reproduced bug reports for Android apps. This dataset was later extended through the addition of another 90 reproduced bug reports in the AndroR2+ dataset~\cite{Johnson:SANER22}, for a total of 180 real, reproducible reports. For each bug report, AndroR2+ provides a link to the original bug report in the issue tracker, an \texttt{\small apk} of the buggy app version, a reproduction script, and metadata for bug reproduction (device, OS version, \etc).

To construct our new dataset of real bugs, we chose the three apps with the largest number of bugs from AndroR2+, while also ensuring the diversity of app categories. We selected: Firefox Focus (FCS)~\cite{fcs}, a web browser; PDF Converter (ITP)~\cite{itp}, an image-to-PDF converter; and GPSTest (GPS)~\cite{gps}, a GPS testing app. FCS is the only app that renders dynamic content on the screen. For these apps in Andror2+, we found ten bug reports for FCS, nine  reports for GPS, and eight reports for ITP. We further manually checked each app's issue tracker and collected one more bug for GPS and two more bugs for ITP to have the same number of bugs per app. To find the \texttt{\small apk} files of the correct buggy version of the apps for these three bugs, we chose the app version closest to the date the issue was created and confirmed that the \texttt{\small apk} allowed for the successful reproduction of the bug. Based on the AndroR2+ metadata and the three bug reports we collected, there are seven different OS versions used to reproduce the bugs, namely, Android version 4.4.4, 6.0.1, 7, 7.1, 8, 8.1, \& 9.

\subsubsection{\textbf{Duplicate Video Recording}}

The paper authors and external participants recorded videos replicating the collected 30 real bugs from the three AndroR2+ apps, following prior work~\cite{cooper2021takes}.  

We rewrote the descriptions of steps to reproduce (S2R), expected behaviors, and observed behaviors for these bugs to ensure they are clear and easy for participants to reproduce from an end-user perspective. Although the AndroR2+ bugs were reproducible on a Pixel 2 emulator, we chose Nexus 5X to maintain the same device configuration as the previous dataset~\cite{cooper2021takes}, since the bugs were also reproducible on the Nexus 5X. This ensures a consistent resolution of the videos across the benchmarks. Additionally, we minimized the different OS versions to three (6, 8.1, and 9) to reduce participants' effort by finding the closest OS versions to their original ones while ensuring the bugs were still reproducible. Also, having these additional OSes in our video reproductions of these bugs has the added benefit of being more realistic---the prior dataset only used Android 7.0. 
While AndroR2+ provides automated bug reproduction scripts, we avoided using them for two reasons: (i) we found that certain scripts led to errors that did not properly reproduce the bug, and (ii) we wanted to capture video-based reports depicting real human actions, to ensure the most realistic setting possible.

Video-based reports were created by the paper authors for all 30 bugs according to the S2R. To maintain three duplicate videos per bug, in line with the previous dataset, two authors (who previously did not record any videos) along with two Ph.D. students were asked to record the additional 60 videos, each responsible for reproducing 15 distinct bugs with only the descriptions of expected and observed behaviors, to ensure diversity of reproduction steps. Unlike the prior dataset, the recorded videos do not show the Android touch indicator when the user taps on the screen. 

In total, our new dataset consists of 90 video-based bug reports corresponding with three duplicates of 30 real bugs from three apps. It contains two crashes and 28 non-crashes, comprising 270 reproduction steps in total (249 taps, six gestures, and 15 input entry actions) and $\approx$35-second videos, on average.
There are six videos for Android 6, nine for Android 9, and 75 for Android 8.1.

\subsubsection{\textbf{Duplicate Detection Tasks}}

In line with the prior dataset, we construct \textit{duplicate detection tasks} for each app to be as realistic as possible. We define a duplicate detection task as having: (1) a \textit{query} video that represents a newly reported video-based bug report, and (2) a \textit{corpus} of 13 existing video-based reports. The query must be compared against the corpus in order to determine whether the incoming report is a duplicate of an existing report. Each task contains videos of the 10 bugs for an app. The corpus contains two duplicate videos of the query (\ie they show the same bug). The remaining eleven videos are non-duplicates: three of them are duplicates of each other but not of the query (\ie they show a bug different from the query bug), and eight videos show distinct bugs. Each task simulates a situation that is similar to crowd-sourced app testing, where duplicates of the query, of other bugs, and unique video-based reports exist together on the issue tracker for an app.

Using different combinations of bugs and videos, we created a total of 810 tasks per app or 2,430 tasks across all apps. Combining both the prior and new datasets, there are 7,920 tasks in our extended evaluation benchmark to evaluate \ap.

\subsection{Baseline Duplicate Detector}

We compare \ap against the \tango duplicate detector introduced by Cooper \etal~\cite{cooper2021takes}. \tango also leverages multi-modal information to detect duplicate video-based reports, using less-sophisticated methods as compared to \ap. It extracts visual features from video frames using a contrastive learning method called SimCLR, which uses a ResNet-50 CNN to learn local features of app GUIs~\cite{chen2020simple}.
It also analyzes text displayed on GUI screens using an approach that combines LSTM-based language models and heuristics, relying on TesseracOCR to extract video frame text~\cite{tesseract-ocr}. 
Finally, \tango performs limited alignment of video frames: only for its extracted visual SimCLR features. 
\tango's evaluation found the best performing configuration is when the visual and textual components are combined, hence we compare \ap against this configuration while also performing ablation comparisons between their individual components.

\subsection{Metrics and Experimental Settings}

\subsubsection{\textbf{Evaluation Metrics}}

We use standard metrics used in prior work on duplicate bug report detection evaluations~\cite{zhang2023duplicate,Lin2016a,Kang2017,cooper2021takes}:
\begin{itemize}[leftmargin=*]
    \item \textbf{Mean Reciprocal Rank (mRR)}: it gives a measure of the average ranking of the first duplicate video found in the candidate list of videos given by a duplicate detector. It is calculated as: $mRR=\frac{1}{N}\sum^{N}_{i=1}\frac{1}{rank_i}$, for $N$ duplicate detection tasks (${rank_i}$ is the rank of the first duplicate video for task $i$).
    
    \item \textbf{Mean Average Precision (\textbf{mAP})}: it gives a measure of the average ranking of all the duplicate videos for a query video. It is computed as: $mAP = \frac{1}{N}\sum^{N}_{i=1}\frac{1}{DV}\sum^{DV}_{v=1}P_i(rank_v)$, where $DV$ is the set of duplicate videos for task $i$, $rank_v$ is the rank of the duplicate video $v$, and $P_i(k)=\frac{duplicates}{k}$ is the number of duplicates in the top-$k$ candidates.
\end{itemize}

All metrics give a normalized score in [0, 1]---the higher the score, the higher the duplicate detection performance. We executed different configurations of \ap and the baseline on the 7,920 tasks and computed/compared the metrics between these approaches.

\subsubsection{\textbf{Model Configurations}}
\label{sec: experiment_config}

We compared \apv against \tango's visual component by experimenting with two ViT models: ViT-Small (ViT-S) and ViT-Base (ViT-B), which have six and 12 self-attention heads respectively.
ViT-S has a similar size to RestNet-50's size (used by \tango's SimCLR): $\approx$23M parameters. To evaluate the differences between the SimCLR (contrastive) and DINO (distillation) training schemes, we implemented \apv with DINO + RestNet-50.
We also experimented with the following patch sizes for ViT: $16 x 16$ (/16) and $8 x 8$ (/8) pixels, as the patch size can affect \apvs performance~\cite{dosovitskiy2020image}. In total, we executed four DINO models: DINO (ResNet), DINO (ViT-S/16), DINO (ViT-S/8), \& DINO (ViT-B/16). ViT-B/8 was not included in the experimentation for \apv due to its substantial computation overhead.

To account for potential biases from random image selection when constructing the \apvs Codebook, we used four distinct Codebooks, each trained on 15k distinct RICO images (60K images in total). With each Codebook, \apv generates similarity scores for a set of videos. These similarities are averaged across the four Codebooks to produce final scores used for ranking. To perform a fair comparison with \tango's visual component, we implemented the same Codebook generation strategy on \tango, using its publicly released implementation~\cite{cooper2021takes}. The recomputed \tango results on the prior dataset are slightly higher than those reported in the original paper (76.2 vs 75.3 mRR and 69.8 vs 67.8 mAP).

We compared \apt against \tango's textual component by experimenting with different configurations for the EAST and TrOCR models. For EAST, we used three different resolution thresholds to filter out small text regions: $5 \times 5$ (EAST-5), $40 \times 20$ (EAST-40), and $80 \times 40$ (EAST-80). The $5 \times 5$ threshold is used by default in EAST. We did not test larger resolutions than $80 \times 40$ to ensure that each textual document created for the video has at least one valid detection. $40 \times 20$ was included as a middle ground to understand the impact of the threshold size on the video similarity calculation. For TrOCR, we used its large version with BEiT Large~\cite{bao2021beit} as the encoder and RoBERTa Large~\cite{liu2019roberta} as the decoder. Two fine-tuned TrOCR-Large models are used, namely TrOCR-p (fine-tuned on the printed text dataset SROIE \cite{huang2019icdar2019}) and  TrOCR-s (finetuned on the synthetic scene text datasets such as ICDAR15 \cite{Karatzas2015ICDAR2C} and SVT \cite{Wang2011EndtoendST}).

\begin{table}[t]
\centering
\small
\caption{The network configurations and fine-tuning hyperparameters for \janusv compared with SimCLR used by \tango}
\label{tabs:model_train}
\begin{tabular}{c|c|c|c|cc}
\toprule
model   & dim   & \# params   &  batch size    & w-temp    & temp  \\  \hline
SimCLR    & 2,048  & 23M   & 1,792    & -- &   --    \\ 
DINO (ResNet)    & 2,048  & 23M   & 96    & 0.03  &   0.03 (0)    \\  
DINO (ViT-S/16)    & 384  & 21M   & 96    & 0.03  &   0.03 (0)    \\ 
DINO (ViT-S/8)    & 384  & 21M   & 18    & 0.04  &   0.05 (30)    \\  
DINO (ViT-B/16)    & 768  & 85M   & 64    & 0.05  &   0.07 (50)    \\  \hline

\end{tabular}
\end{table}

\subsubsection{\textbf{Model Training}}

All visual models were fine-tuned on the 66k mobile app screenshots from the RICO dataset~\cite{deka2017rico} for 100 epochs using model checkpoints trained on ImageNet~\cite{deng2009imagenet}, except for DINO (ViT-B/16), to fairly compare it with the \tangovs SimCLR model. After examining preliminary results showing the advantages of DINO with ViT, we decided to train DINO (ViT-B/16) for 400 epochs~\cite{caron2021emerging}. Fine-tuning was carried out on three NVIDIA T4 Tesla GPUs with 16GB of memory each. Because DINO does not use contrastive learning, we were able to use a much smaller batch size as compared to the SimCLR model used in \tango: 96 \textit{vs} 1,792 for ViT-S/16 and ResNet-50. For the ViT-B/16 and ViT-S/8 models, we used a batch size of 64 and 16 due to memory constraints. Table \ref{tabs:model_train} shows the network configurations and three fine-tuning hyperparameters, where \textit{dim} is the representation dimension of the output, \textit{\# params} is the total number of model parameters. \textit{"temp" and "w-temp"} represent the teacher temperature and the warm-up teacher temperature respectively, and the numbers in parentheses are the \# epochs used for warm-up. Model training was not required for \janust as we directly use pre-trained EAST and TrCOR models for GUI text localization and recognition~\cite{zhou2017east,li2021trocr}.
\section{Evaluation Results and Discussion}\label{sec:results}

\Cref{tab:individual_results} shows \aps duplicate detection performance compared to the baseline \tango, for their individual components: visual, textual, and sequential. \Cref{tab:combined_results} shows the performance of different combinations of \ap components, compared to the baseline.

Cells shaded green in these tables indicate a statistically significant (via Wilcoxon's paired test at the $p<0.05$ level) higher effectiveness when comparing a given \ap configuration/component to a given \tango configuration/component. Yellow shaded cells indicate a higher performance, but without statistical significance. We present the results for each app of the \textit{original} (mostly synthetic bugs) and \textit{extended} (real bugs) \textit{datasets} and the overall results accounting for all the apps in both sets, separately and combined. 

While we computed the performance of four \janusv DINO models (\ie DINO with ResNet, ViT-S/16, ViT-S/8, and ViT-B/16), we present (in \cref{tab:individual_results,tab:combined_results}) the best-performing model for \janusv: DINO with ViT-B. Likewise, we report here the results of the best performing model configuration for \apt, namely EAST-80 (EAST that filters out region proposals smaller than $80 x 40$) combined with TrOCR-s (TrOCR fine-tuned on real-world scenes, \eg street scenes, instead of text found in printed and handwritten documents). The results for all the DINO, EAST, and TrOCR configurations can be found in our replication package~\cite{janus-tool}.

\Cref{tab:individual_results,tab:combined_results} show a consistent trend: the performance achieved by any duplicate detector (\ie any configuration) is lower for the original dataset than for the extended dataset. After investigating the minimal set of ground-truth reproduction steps of the bugs used in the datasets, we found this trend is explained by the number of overlapping steps across distinct bugs in an app. We observed that distinct bugs for a given app in the original dataset have a larger step overlap than distinct bugs in the extended dataset. It is more challenging for a duplicate detector to distinguish between duplicate and non-duplicate videos if there is a larger step overlap across bugs (hence, across videos). Recall that in a duplicate detection task, the videos in the corpus are for distinct bugs; if there is a larger overlap among them, particularly between duplicates and non-duplicates, a detector would struggle to discern the differences. 

\subsection{RQ1: \apvs Performance}

\begin{table*}[]
\caption{Performance of the individual components of \ap and the baseline \tango}
\label{tab:individual_results}
\begin{adjustbox}{width=0.9\linewidth,center}
\begin{tabular}{c?cccc?cccc?cccc?cc}
\hline
                               & \multicolumn{4}{c?}{\textbf{Visual}}                                                                                                         & \multicolumn{4}{c?}{\textbf{Textual}}                                                                                                        & \multicolumn{4}{c?}{\textbf{Sequential (visual)}}                                                                                 & \multicolumn{2}{l}{\textbf{Seq. (textual)}}     \\ \cline{2-15} 
                               & \multicolumn{2}{c|}{\textbf{\tango}}                          & \multicolumn{2}{c?}{\textbf{\apv}}             & \multicolumn{2}{c|}{\textbf{\tango}}                          & \multicolumn{2}{c?}{\textbf{\apt}}             & \multicolumn{2}{c|}{\textbf{\tango}}                          & \multicolumn{2}{c?}{\textbf{\apsv}} & \multicolumn{2}{c}{\textbf{\apst}}     \\ \cline{2-15} 
\multirow{-3}{*}{\textbf{App}} & \textbf{mRR}   & \multicolumn{1}{c|}{\textbf{mAP}}                           & \textbf{mRR}                  & \textbf{mAP}                  & \textbf{mRR}   & \multicolumn{1}{c|}{\textbf{mAP}}                           & \textbf{mRR}                  & \textbf{mAP}                  & \textbf{mRR}   & \multicolumn{1}{c|}{\textbf{mAP}}                           & \textbf{mRR}             & \textbf{mAP}            & \textbf{mRR}   & \textbf{mAP} \\ \hline
APOD                  & 77.19          & \multicolumn{1}{c|}{\cellcolor[HTML]{FFFFFF}69.98}          & \multicolumn{1}{c}{\cellcolor[HTML]{B6D7A8}87.32}                         & \multicolumn{1}{c?}{\cellcolor[HTML]{B6D7A8}79.79}                         & \cellcolor[HTML]{FFE599}80.80          & \multicolumn{1}{c|}{\cellcolor[HTML]{FFFFFF}75.30}          & \cellcolor[HTML]{FFFFFF}79.76 & \cellcolor[HTML]{B6D7A8}73.09 & 55.01          & \multicolumn{1}{c|}{\cellcolor[HTML]{FFFFFF}44.85}          & \multicolumn{1}{c}{\cellcolor[HTML]{B6D7A8}84.45}                    & \multicolumn{1}{c?}{\cellcolor[HTML]{B6D7A8}71.11}                   & 73.40          & 68.09                                \\
\rowcolor[HTML]{FFFFFF} 
DROID               & 68.43          & \multicolumn{1}{c|}{\cellcolor[HTML]{FFFFFF}58.82}          & \multicolumn{1}{c}{\cellcolor[HTML]{B6D7A8}80.77}                         & \multicolumn{1}{c?}{\cellcolor[HTML]{B6D7A8}71.44}                         & 67.90          & \multicolumn{1}{c|}{\cellcolor[HTML]{FFFFFF}64.70}          & \multicolumn{1}{c}{\cellcolor[HTML]{B6D7A8}78.88}                         & \multicolumn{1}{c?}{\cellcolor[HTML]{B6D7A8}72.52}                         & 46.54          & \multicolumn{1}{c|}{\cellcolor[HTML]{FFFFFF}37.91}          & \multicolumn{1}{c}{\cellcolor[HTML]{B6D7A8}61.49}                    & \multicolumn{1}{c?}{\cellcolor[HTML]{B6D7A8}50.88}                   & 76.19          & 73.64                                \\
\rowcolor[HTML]{FFFFFF} 
GNU                   & 81.53          & \multicolumn{1}{c|}{\cellcolor[HTML]{FFE599}75.83}          & \cellcolor[HTML]{FFE599}81.83 & \cellcolor[HTML]{FFFFFF}75.54 & \cellcolor[HTML]{FFE599}84.50          & \multicolumn{1}{c|}{\cellcolor[HTML]{B6D7A8}82.30}          & \multicolumn{1}{c}{\cellcolor[HTML]{FFFFFF}89.53}                         & \cellcolor[HTML]{FFFFFF}81.28 & 55.91          & \multicolumn{1}{c|}{\cellcolor[HTML]{FFFFFF}43.37}          & \multicolumn{1}{c}{\cellcolor[HTML]{B6D7A8}71.41}                    & \multicolumn{1}{c?}{\cellcolor[HTML]{B6D7A8}58.82}                   & 52.79          & 43.81                                \\
\rowcolor[HTML]{FFFFFF} 
GROW                  & 83.53          & \multicolumn{1}{c|}{\cellcolor[HTML]{FFFFFF}78.60}          & \multicolumn{1}{c}{\cellcolor[HTML]{B6D7A8}87.46}                         & \multicolumn{1}{c?}{\cellcolor[HTML]{B6D7A8}80.33}                         & 76.80          & \multicolumn{1}{c|}{\cellcolor[HTML]{FFFFFF}69.00}          & \multicolumn{1}{c}{\cellcolor[HTML]{B6D7A8}82.65}                         & \multicolumn{1}{c?}{\cellcolor[HTML]{B6D7A8}77.38}                         & 74.57          & \multicolumn{1}{c|}{\cellcolor[HTML]{FFFFFF}64.46}          & \multicolumn{1}{c}{\cellcolor[HTML]{B6D7A8}92.14}                    & \multicolumn{1}{c?}{\cellcolor[HTML]{B6D7A8}84.57}                   & 77.45          & 76.24                                \\
\rowcolor[HTML]{FFFFFF} 
TIME                  & 70.26          & \multicolumn{1}{c|}{\cellcolor[HTML]{FFFFFF}65.35}          & \multicolumn{1}{c}{\cellcolor[HTML]{B6D7A8}73.76}                         & \multicolumn{1}{c?}{\cellcolor[HTML]{B6D7A8}69.46}                         & 47.40          & \multicolumn{1}{c|}{\cellcolor[HTML]{FFFFFF}37.70}          & \multicolumn{1}{c}{\cellcolor[HTML]{B6D7A8}64.80}                         & \multicolumn{1}{c?}{\cellcolor[HTML]{B6D7A8}56.67}                         & 50.85          & \multicolumn{1}{c|}{\cellcolor[HTML]{FFFFFF}43.62}          & \multicolumn{1}{c}{\cellcolor[HTML]{B6D7A8}63.14}                    & \multicolumn{1}{c?}{\cellcolor[HTML]{B6D7A8}56.18}                   & 69.34          & 66.16                                \\
\rowcolor[HTML]{FFFFFF} 
TOK                   & 76.03          & \multicolumn{1}{c|}{\cellcolor[HTML]{FFFFFF}70.37}          & \multicolumn{1}{c}{\cellcolor[HTML]{B6D7A8}81.11}                         & \cellcolor[HTML]{FFE599}71.33 & \multicolumn{1}{c}{\cellcolor[HTML]{B6D7A8}61.30}          & \multicolumn{1}{c|}{\cellcolor[HTML]{B6D7A8}53.30}          & \cellcolor[HTML]{FFFFFF}53.95 & \cellcolor[HTML]{FFFFFF}44.48 & 38.13          & \multicolumn{1}{c|}{\cellcolor[HTML]{FFFFFF}33.36}          & \multicolumn{1}{c}{\cellcolor[HTML]{B6D7A8}53.39}                    & \multicolumn{1}{c?}{\cellcolor[HTML]{B6D7A8}43.22}                   & 54.00          & 47.83                                \\ \hline
\rowcolor[HTML]{FFFFFF} 
\textbf{Original}       & \textbf{76.16} & \multicolumn{1}{c|}{\cellcolor[HTML]{FFFFFF}\textbf{69.83}} & \multicolumn{1}{c}{\cellcolor[HTML]{B6D7A8}\textbf{82.04}}                & \multicolumn{1}{c?}{\cellcolor[HTML]{B6D7A8}\textbf{74.65}}                & \textbf{69.80} & \multicolumn{1}{c|}{\cellcolor[HTML]{FFFFFF}\textbf{63.70}} & \multicolumn{1}{c}{\cellcolor[HTML]{B6D7A8}\textbf{74.93}}                & \multicolumn{1}{c?}{\cellcolor[HTML]{B6D7A8}\textbf{67.57}}                & \textbf{53.50} & \multicolumn{1}{c|}{\cellcolor[HTML]{FFFFFF}\textbf{44.59}} & \multicolumn{1}{c}{\cellcolor[HTML]{B6D7A8}\textbf{71.00}}           & \multicolumn{1}{c?}{\cellcolor[HTML]{B6D7A8}\textbf{60.80}}          & \textbf{67.20} & \textbf{62.63}                       \\ \hline
\rowcolor[HTML]{FFFFFF} 
FCS                   & \multicolumn{1}{c}{\cellcolor[HTML]{B6D7A8}91.09}          & \multicolumn{1}{c|}{\cellcolor[HTML]{B6D7A8}85.82}          & 86.88 & 82.69 & 85.12          & \multicolumn{1}{c|}{\cellcolor[HTML]{FFFFFF}79.12}          & \cellcolor[HTML]{FFE599}86.17 & \multicolumn{1}{c?}{\cellcolor[HTML]{B6D7A8}84.88}                         & 65.23          & \multicolumn{1}{c|}{\cellcolor[HTML]{FFFFFF}55.42}          & \multicolumn{1}{c}{\cellcolor[HTML]{B6D7A8}90.20}                    & \multicolumn{1}{c?}{\cellcolor[HTML]{B6D7A8}85.91}                   & 90.53          & 88.21                                \\
\rowcolor[HTML]{FFFFFF} 
GPS                   & 95.99          & \multicolumn{1}{c|}{\cellcolor[HTML]{FFFFFF}92.15}          & \multicolumn{1}{c}{\cellcolor[HTML]{B6D7A8}98.09}                         & \multicolumn{1}{c?}{\cellcolor[HTML]{B6D7A8}95.70}                         & 92.11          & \multicolumn{1}{c|}{\cellcolor[HTML]{FFFFFF}84.82}          & \multicolumn{1}{c}{\cellcolor[HTML]{B6D7A8}97.51}                         & \multicolumn{1}{c?}{\cellcolor[HTML]{B6D7A8}96.10}                         & 68.34          & \multicolumn{1}{c|}{\cellcolor[HTML]{FFFFFF}60.63}          & \multicolumn{1}{c}{\cellcolor[HTML]{B6D7A8}93.72}                    & \multicolumn{1}{c?}{\cellcolor[HTML]{B6D7A8}86.64}                   & 57.83          & 53.33                                \\
\rowcolor[HTML]{FFFFFF} 
ITP                   & 81.93          & \multicolumn{1}{c|}{\cellcolor[HTML]{FFFFFF}73.92}          & \multicolumn{1}{c}{\cellcolor[HTML]{B6D7A8}93.29}                         & \multicolumn{1}{c?}{\cellcolor[HTML]{B6D7A8}84.08}                         & 89.73          & \multicolumn{1}{c|}{\cellcolor[HTML]{FFFFFF}86.34}          & \multicolumn{1}{c}{\cellcolor[HTML]{B6D7A8}96.77}                         & \multicolumn{1}{c?}{\cellcolor[HTML]{B6D7A8}89.83}                         & 69.50          & \multicolumn{1}{c|}{\cellcolor[HTML]{FFFFFF}54.37}          & \multicolumn{1}{c}{\cellcolor[HTML]{B6D7A8}90.56}                    & \multicolumn{1}{c?}{\cellcolor[HTML]{B6D7A8}78.20}                   & 54.74          & 46.87                                \\ \hline
\rowcolor[HTML]{FFFFFF} 
\textbf{Extended}        & \textbf{89.67} & \multicolumn{1}{c|}{\cellcolor[HTML]{FFFFFF}\textbf{83.96}} & \multicolumn{1}{c}{\cellcolor[HTML]{B6D7A8}\textbf{92.75}}                & \multicolumn{1}{c?}{\cellcolor[HTML]{B6D7A8}\textbf{87.49}}                & \textbf{88.98} & \multicolumn{1}{c|}{\cellcolor[HTML]{FFFFFF}\textbf{85.74}} & \multicolumn{1}{c}{\cellcolor[HTML]{B6D7A8}\textbf{93.48}}                & \multicolumn{1}{c?}{\cellcolor[HTML]{B6D7A8}\textbf{90.27}}                & \textbf{67.69} & \multicolumn{1}{c|}{\cellcolor[HTML]{FFFFFF}\textbf{56.81}} & \multicolumn{1}{c}{\cellcolor[HTML]{B6D7A8}\textbf{91.50}}           & \multicolumn{1}{c?}{\cellcolor[HTML]{B6D7A8}\textbf{83.58}}          & \textbf{67.70} & \textbf{62.80}                       \\ \hline
\rowcolor[HTML]{FFFFFF} 
\textbf{Overall}               & \textbf{80.66} & \multicolumn{1}{c|}{\cellcolor[HTML]{FFFFFF}\textbf{74.54}} & \multicolumn{1}{c}{\cellcolor[HTML]{B6D7A8}\textbf{85.61}}                & \multicolumn{1}{c?}{\cellcolor[HTML]{B6D7A8}\textbf{78.93}}                & \textbf{76.14} & \multicolumn{1}{c|}{\cellcolor[HTML]{FFFFFF}\textbf{70.06}} & \multicolumn{1}{c}{\cellcolor[HTML]{B6D7A8}\textbf{81.11}}                & \multicolumn{1}{c?}{\cellcolor[HTML]{B6D7A8}\textbf{75.14}}                & \textbf{58.23} & \multicolumn{1}{c|}{\cellcolor[HTML]{FFFFFF}\textbf{48.67}} & \multicolumn{1}{c}{\cellcolor[HTML]{B6D7A8}\textbf{77.84}}           & \multicolumn{1}{c?}{\cellcolor[HTML]{B6D7A8}\textbf{68.39}}          & \textbf{67.36} & \textbf{62.69}                       \\ \hline
\end{tabular}
\end{adjustbox}
\end{table*}

\Cref {tab:individual_results} shows the duplicate detection effectiveness of \janusv (DINO with ViT-B) compared to visual \tango (SimCLR). 

Before discussing the table results, we briefly discuss the results of comparing the training schemes (distillation via DINO \textit{vs} contrastive via SimCLR, both using the same pre-trained ResNet weights). We found that SimCLR outperforms DINO for six of nine apps by a relatively small margin (by 3.5\% mRR and 4.2\% mAP, on average), but DINO outperforms SimCLR for the remaining three apps (APOD, GNU, DINO) by a larger margin (7.5\% mRR and 6\% mAP, on avg.). Overall, across all the apps, we found a similar performance between these two approaches (less than 1.1\% mRR/mAP improvement), which indicates the training scheme does not have a large impact on duplicate detection performance. 

Furthermore, both ViT-S/16 and ViT-S/8 used by \janusvs DINO exhibit superior performance compared to ResNet-50 used by visual \tangos SimCLR. Specifically, although ViT-S/16 and ViT-S/8 have a similar model size to RestNet-50, they outperform ResNet-50 by 2.91\% and 2.92\% respectively, in terms of mRR on average, with statistical significance. This highlights the effectiveness of ViT  over ResNet for duplicate video-based bug report detection.

\Cref {tab:individual_results} shows that \janusv (DINO with ViT-B) significantly outperforms the baseline on both datasets. We observe an overall improvement of (85.61 - 80.66)/80.66 = 6.1\% mRR and (78.93 - 74.54)/74.54 = 5.9\% mAP,  with statistical significance. This overall performance comes from an improvement in eight out of nine apps compared to the baseline (seven with statistical significance), with \tango only having a substantial improvement over \apv for the FCS app. These FCS results are due to the nature of the app and the underlying models. Specifically, FCS is a web browser and the video-based bug reports produced for this app show users navigating to different websites. The app produces dynamic content: the navigated websites have different layouts and visual characteristics. \apvs ViT is prone to focusing more on the structure of the GUIs, extracting global features about the layouts, while the baseline's ResNet tends to focus on local visual features of the GUIs, not necessarily on general screen layouts, which are more beneficial to detect duplicates. Compared to ResNet, ViT's emphasis on GUI layouts leads to a more substantial dissimilarity between duplicates when sequential visual information is not taken into account.

\subsection{RQ2: \apts Performance}   

\Cref{tab:individual_results} shows that \apt is substantially more effective than textual \tango for seven of nine apps (with statistical significance), especially for DROID (improvement of 16.2\% mRR and 12.1\% mAP) and TIME (improvement of 36.7\% mRR and 50.3\% mAP). Only for the apps APOD and TOK, \tango is higher, resulting in \apt's overall superiority on both the original and the real bug datasets (overall, by 6.5\%/7.3\% mRR/mAP). 
The reason why \apt does not perform better for APOD and TOK is that these apps usually contain short or small pieces of text (\eg due to small fonts) on many of their screens, and EAST fails to identify them because these pieces fit in smaller regions than $80 \times 40$ pixels.
Indeed, when reducing the threshold to $40 \times 20$, \apt outperforms \tango for APOD and TOK (by 1.5\%/1.1\% and 8.8\%/5.2\% mRR/mAP respectively).  

\apts performance is slightly higher than \apvs for the extended dataset (improvement of 0.8\%/3.1\% mRR/mAP overall), but lower for the original dataset (by 9.5\%/10.5\% mRR/mAP). The lower improvements come from the TOK app, which does not contain enough textual information to accurately 
detect duplicates~\cite{cooper2021takes}. 

\subsection{RQ3: \apsss Performance}

\Cref{tab:individual_results} shows that  \apsv is substantially more effective in detecting duplicates than sequential \tango, when using visual frame representations. \apsv outperforms the baseline for every app in the original dataset (by 32.7\% mRR and 36.4\% mAP overall) and in the extended dataset (by 35.2\% mRR and 47.1\% mAP overall). The high improvements can be attributed to the power of \aps ViT in learning the global structure of GUI screens, while \tango's ResNet focuses on learning local GUI features.  Global GUI structure representations are more useful to measure the sequential overlap between video frames even when there are small variations in the frames because of slightly different reproduction steps. Also, we note that the improvements for the FCS app are substantial (38\%/55\% mRR/mAP). Since we observed highly different GUI layouts in video frames for this app (because users navigated to different websites), these results are indicative of the effectiveness of the sequential similarity approach of \ap in combination with ViT-based frame representations (compared to the baseline).

Since \tango's sequential component is not designed to work with textual frame representations (unlike \ap), we only compare the performance of \apst with \apsv. Overall, \apsv outperforms \apst by 15.6\%/9.1\% mRR/mAP. It substantially outperforms \apst for five of nine apps by 39.4\%/ 35.8\% mRR/mAP on average, while having a lower performance for the remaining four apps (7.4\%/14.6\% mRR/mAP). The largest improvement is observed in the GPS and ITP apps. ITP is an app used to convert images to PDF, involving mostly image editing, while GPS focuses on editing coordinates and displaying locations on a map. Consequently, video-based bug reports have limited text on each frame, which negatively impacts the performance of \apst. However, since \apt leads to high performance for these two apps, we attribute \apsts relatively low performance to the alignment approach, which processes each video frame text rather than using the text from all frames together.

\begin{table}[]
\caption{Performance of different component combinations for  \ap and the baseline \tango}
\setlength{\tabcolsep}{2pt}
\label{tab:combined_results}
\begin{adjustbox}{width=\linewidth,center}
\begin{tabular}{c|cccc|cc|cc|cc}
\hline
                               & \multicolumn{4}{c|}{\textbf{Visual + Textual}}                                                                                                                                         & \multicolumn{2}{c|}{\textbf{Vis + Seq}}                                     & \multicolumn{2}{c|}{\textbf{Txt + Seq}}                                    & \multicolumn{2}{c}{\textbf{Vis+Txt+Seq}}                                 \\ \cline{2-11} 
                               & \multicolumn{2}{c|}{\textbf{\tango}}                                          & \multicolumn{2}{c|}{\textbf{\ap}}                     & \multicolumn{2}{c|}{\textbf{\ap}}                     & \multicolumn{2}{c|}{\textbf{\ap}}                     & \multicolumn{2}{c}{\textbf{\ap}}                      \\ \cline{2-11} 
\multirow{-3}{*}{\textbf{App}} & \textbf{mRR}   & \multicolumn{1}{c|}{\textbf{mAP}}   & \textbf{mRR}   & \textbf{mAP}   & \textbf{mRR}   & \textbf{mAP}   & \textbf{mRR}   & \textbf{mAP}   & \textbf{mRR}   & \textbf{mAP}   \\ \hline
APOD                           & \cellcolor[HTML]{FFFFFF}81.08          & \multicolumn{1}{c|}{\cellcolor[HTML]{FFFFFF}75.11}          & \cellcolor[HTML]{B6D7A8}86.72          & \cellcolor[HTML]{B6D7A8}80.64          & \cellcolor[HTML]{B6D7A8}86.59          & \cellcolor[HTML]{FFE599}77.30          & \cellcolor[HTML]{B6D7A8}91.54          & \cellcolor[HTML]{B6D7A8}86.30          & \cellcolor[HTML]{B6D7A8}94.95          & \cellcolor[HTML]{B6D7A8}86.55          \\
DROID                          & \cellcolor[HTML]{FFFFFF}71.74          & \multicolumn{1}{c|}{\cellcolor[HTML]{FFFFFF}64.31}          & \cellcolor[HTML]{B6D7A8}83.95          & \cellcolor[HTML]{B6D7A8}78.98          & \cellcolor[HTML]{B6D7A8}75.50          & \cellcolor[HTML]{FFFFFF}64.96          & \cellcolor[HTML]{B6D7A8}89.26          & \cellcolor[HTML]{B6D7A8}83.15          & \cellcolor[HTML]{B6D7A8}88.56          & \cellcolor[HTML]{B6D7A8}81.06          \\
GNU                            & \cellcolor[HTML]{FFFFFF}89.16          & \multicolumn{1}{c|}{\cellcolor[HTML]{FFFFFF}85.92}          & \cellcolor[HTML]{FFE599}89.62          & \cellcolor[HTML]{FFFFFF}81.75          & \cellcolor[HTML]{FFFFFF}83.48          & \cellcolor[HTML]{FFFFFF}74.18          & \cellcolor[HTML]{FFFFFF}84.24          & \cellcolor[HTML]{FFFFFF}73.80          & \cellcolor[HTML]{FFE599}90.58          & \cellcolor[HTML]{FFFFFF}81.58          \\
GROW                           & \cellcolor[HTML]{FFFFFF}86.61          & \multicolumn{1}{c|}{\cellcolor[HTML]{FFFFFF}80.73}          & \cellcolor[HTML]{B6D7A8}89.84          & \cellcolor[HTML]{B6D7A8}86.32          & \cellcolor[HTML]{B6D7A8}91.40          & \cellcolor[HTML]{B6D7A8}86.88          & 83.56          & \cellcolor[HTML]{B6D7A8}80.96          & \cellcolor[HTML]{B6D7A8}93.32          & \cellcolor[HTML]{B6D7A8}90.72          \\
TIME                           & \cellcolor[HTML]{FFFFFF}65.06          & \multicolumn{1}{c|}{\cellcolor[HTML]{FFFFFF}59.23}          & \cellcolor[HTML]{B6D7A8}67.51          & \cellcolor[HTML]{B6D7A8}64.31          & \cellcolor[HTML]{B6D7A8}71.33          & \cellcolor[HTML]{B6D7A8}63.68          & \cellcolor[HTML]{B6D7A8}73.69          & \cellcolor[HTML]{B6D7A8}69.67          & \cellcolor[HTML]{B6D7A8}74.88          & \cellcolor[HTML]{B6D7A8}71.92          \\
TOK                            & \cellcolor[HTML]{FFFFFF}71.11          & \multicolumn{1}{c|}{\cellcolor[HTML]{FFFFFF}63.95}          & \cellcolor[HTML]{B6D7A8}75.51          & 62.59          & \cellcolor[HTML]{FFFFFF}63.35          & \cellcolor[HTML]{FFFFFF}57.30          & \cellcolor[HTML]{FFFFFF}55.23          & \cellcolor[HTML]{FFFFFF}48.57          & \cellcolor[HTML]{B6D7A8}75.92          & \cellcolor[HTML]{B6D7A8}67.91          \\ \hline
\textbf{Orig.}                 & \cellcolor[HTML]{FFFFFF}\textbf{77.46} & \multicolumn{1}{c|}{\cellcolor[HTML]{FFFFFF}\textbf{71.54}} & \cellcolor[HTML]{B6D7A8}\textbf{82.19} & \cellcolor[HTML]{B6D7A8}\textbf{75.76} & \cellcolor[HTML]{B6D7A8}\textbf{78.61} & \textbf{70.72} & \cellcolor[HTML]{B6D7A8}\textbf{79.59} & \cellcolor[HTML]{B6D7A8}\textbf{73.74} & \cellcolor[HTML]{B6D7A8}\textbf{86.37} & \cellcolor[HTML]{B6D7A8}\textbf{79.95} \\ \hline
FCS                            & \cellcolor[HTML]{FFFFFF}91.11          & \multicolumn{1}{c|}{\cellcolor[HTML]{FFFFFF}86.87}          & \cellcolor[HTML]{FFFFFF}88.46          & \cellcolor[HTML]{FFFFFF}85.95          & \cellcolor[HTML]{B6D7A8}93.98          & \cellcolor[HTML]{B6D7A8}89.85          & 90.38          & 84.74          & \cellcolor[HTML]{B6D7A8}94.73          & \cellcolor[HTML]{B6D7A8}91.90          \\
GPS                            & \cellcolor[HTML]{FFFFFF}97.35          & \multicolumn{1}{c|}{\cellcolor[HTML]{FFFFFF}95.53}          & \cellcolor[HTML]{B6D7A8}99.30          & \cellcolor[HTML]{B6D7A8}98.31          & \cellcolor[HTML]{B6D7A8}98.09          & \cellcolor[HTML]{FFE599}96.01          & \cellcolor[HTML]{FFFFFF}89.20          & \cellcolor[HTML]{FFFFFF}86.91          & \cellcolor[HTML]{B6D7A8}98.24          & \cellcolor[HTML]{B6D7A8}97.26          \\
ITP                            & \cellcolor[HTML]{FFFFFF}90.64          & \multicolumn{1}{c|}{\cellcolor[HTML]{FFFFFF}86.51}          & \cellcolor[HTML]{B6D7A8}96.84          & \cellcolor[HTML]{B6D7A8}91.58          & \cellcolor[HTML]{B6D7A8}96.05          & \cellcolor[HTML]{B6D7A8}89.99          & \cellcolor[HTML]{FFFFFF}83.94          & \cellcolor[HTML]{FFFFFF}77.07          & \cellcolor[HTML]{B6D7A8}97.41          & \cellcolor[HTML]{B6D7A8}93.51          \\ \hline
\textbf{Ext.}                  & \cellcolor[HTML]{FFFFFF}\textbf{93.03} & \multicolumn{1}{c|}{\cellcolor[HTML]{FFFFFF}\textbf{89.64}} & \cellcolor[HTML]{B6D7A8}\textbf{94.87} & \cellcolor[HTML]{B6D7A8}\textbf{91.94} & \cellcolor[HTML]{B6D7A8}\textbf{96.04} & \cellcolor[HTML]{B6D7A8}\textbf{91.95} & \cellcolor[HTML]{FFFFFF}\textbf{87.84} & \cellcolor[HTML]{FFFFFF}\textbf{82.91} & \cellcolor[HTML]{B6D7A8}\textbf{96.79} & \cellcolor[HTML]{B6D7A8}\textbf{94.22} \\ \hline
\textbf{Overall}               & \cellcolor[HTML]{FFFFFF}\textbf{82.65} & \multicolumn{1}{c|}{\cellcolor[HTML]{FFFFFF}\textbf{77.57}} & \cellcolor[HTML]{B6D7A8}\textbf{86.42} & \cellcolor[HTML]{B6D7A8}\textbf{81.16} & \cellcolor[HTML]{B6D7A8}\textbf{84.42} & \cellcolor[HTML]{B6D7A8}\textbf{77.79} & \textbf{82.34} & \textbf{76.80} & \cellcolor[HTML]{B6D7A8}\textbf{89.84} & \cellcolor[HTML]{B6D7A8}\textbf{84.71} \\ \hline
\end{tabular}
\end{adjustbox}
\end{table}

\subsection{RQ4: Component Combination Performance}  

We linearly combined \aps components (as described in \cref{sec:approach_combinations}) to determine how much they improve the performance, compared to the baseline and individual components. We experimented with different weights (from 0 to 1 in 0.1 increments) using all duplicate detection tasks and selected the weights that lead to the highest mRR/mAP performance.

As mentioned earlier, the best \tango configuration is when its visual and textual components are combined (with a weight of 0.8 and 0.2, respectively), as reported in the original paper~\cite{cooper2021takes}. \aps visual and textual components (\ie~\apv and \apt) are combined using 0.9 and 0.1 as weights. This combination is denoted as "Visual + Textual" in Table~\ref{tab:combined_results}. The table also shows the combination of \aps visual/textual components and the sequential one: ``Vis + Seq'' denotes the average of the similarity scores produced by \janusv and \apsv, while ``Txt + Seq'' denotes the average of the similarity scores produced by \apt and \apst. An average combination means a weight of 0.5.
Finally, we combine the similarities produced by the last two combinations using a weighted linear combination as follows: Sim(Vis + Seq) $\times$ 0.6 + Sim(Txt + Seq) $\times$ 0.4. This combination incorporates every information source from the videos and is denoted as ``Vis + Txt + Seq''.

\Cref{tab:combined_results} shows that the best performing \ap combinations are ``Visual + Textual'' and ``Vis + Txt + Seq'', both outperforming the baseline by 4.6\%/4.6\% mRR/mAP and 8.7\%/9.2\% mRR/mAP overall respectively (with statistical significance). The other two \ap combinations lead to mixed results: ``Vis + Seq'' leads to overall performance gains 
while ``Txt + Seq'' does not produce overall gains, due to its lower performance on the extended dataset. 
 
When using ``Visual + Textual'', \ap significantly outperforms \tango on seven of nine apps and is only worse than \tango on FCS, considering both mRR and mAP. As previously mentioned, \aps lower performance for FCS, compared to \tango, stems from the nature of the app itself. FCS is a web browser and the bugs used for this app were not dependent on a particular web page. When reproducing the bugs, the users navigated to different web pages, each one having different layouts and appearances. This means that the duplicate video-based bug reports appeared to be substantially different. Since \ap focuses more heavily on global GUI layout information, via its DINO+ViT model, \ap struggles to differentiate duplicates from non-duplicates. The local features learned by \tango seem to be useful for duplicate detection even when the duplicate videos show different layouts. The lower \ap mAP value on GNU is explained by the lower mAP values of \apv and \apt on that app (by 0.4\% and 1.2\%---see Table~\ref{tab:individual_results}). 

\aps configuration ``Vis + Txt + Seq'' consistently shows mRR/mAP improvement in all nine apps except GNU, when compared to the baseline \tango. Across these apps, we observe improvements ranging from 6.8\%/6.2\% to 23.4\%/26\% mRR/MAP in the original dataset, and from 0.9\%/1.8\% to 7.5\%/8.1\% mRR/MAP in the extended dataset. This is interesting because the performance of the individual components of this configuration is substantially different across the apps. For instance, for TOK, the sequential aspect of the videos, individually combined with \apv or \apt, is less effective than \tango, but when \apv and \apt are combined together with \apss, \ap leads to substantial improvement (by 6.8\%/6.2\% mRR/mAP). Another example is the FCS app, which seems to benefit from the visual and sequential information, as \apv+\apsv seems to contribute most to the overall performance of the `` Vis + Txt + Seq'' configuration. This suggests that the incorporation of sequential information enhances the \aps ability to handle dynamic content, resulting in improved performance in comparison to its ``Visual + Textual''  configuration and the baseline \tango.

\begin{tcolorbox}[boxsep=1pt,left=2pt,right=2pt,top=0pt,bottom=0pt]
\small
\textbf{Best \ap configuration:} The best performing \ap configuration is when combining visual (\apv), textual (\apt), and sequential information (\apss) from video-based bug reports. This configuration consistently outperforms the baseline duplicate detector for 8/9 mobile apps. It achieves an overall performance of 89.8\%/84.7 mRR/mAP, outperforming the baseline by 8.7\%/9.2\% mRR/mAP. This means that \ap can reduce the effort that developers spend determining if a new video-based bug report shows a known bug (by $(1.60-1.38)/1.38=16\%$, based on avg. rank), since they would need to inspect only 1.38 videos on average (\ie 1.38 avg. rank across all tasks) for finding the first duplicate video in the candidate duplicates suggested by \ap.
\end{tcolorbox}

\subsection{Qualitative Analysis}

We discuss two qualitative examples that illustrate the validity of our hypothesis that the richer representations learned by \aps transformer-based visual representation and OCR models improve duplicate detection for video-based bug reports.

\subsubsection{\textbf{Example 1: Transformer-based Representations Capture Subtle GUI patterns}}
To illustrate why we observed improvement in visual \approach as compared to visual \tango, we use interpretability techniques that generate saliency maps that help visualize the learned visual features. To visualize patterns learned by CNNs, we use a technique called AGF~\cite{Gur2021}. Although AGF can visualize self-supervised models such as SimCLR (used by the baseline), this requires training a supervised linear classifier after each layer and a dedicated algorithm to extract the segmentation information from their weights. Therefore, to simplify our comparison, instead of visualizing SimCLR directly, we visualize its main component, the ResNet-50 CNN using AGF under supervision. We follow past work and use the pre-trained ResNet-50 (on ImageNet~\cite{deng2009imagenet}: the training dataset for ResNet) to generate the saliency map based on the  class IDs with the highest probabilities for a given target GUI screen~\cite{Gur2021}. 
We further visualized the ViT-S/16 model (used by \apvs DINO) by directly displaying the self-attention maps. Visualization of ViTs does not necessitate sophisticated algorithms, given the inherent attention mechanism within these architectures.

\begin{figure}[t] 
\centering 
\includegraphics[width=0.45\textwidth]{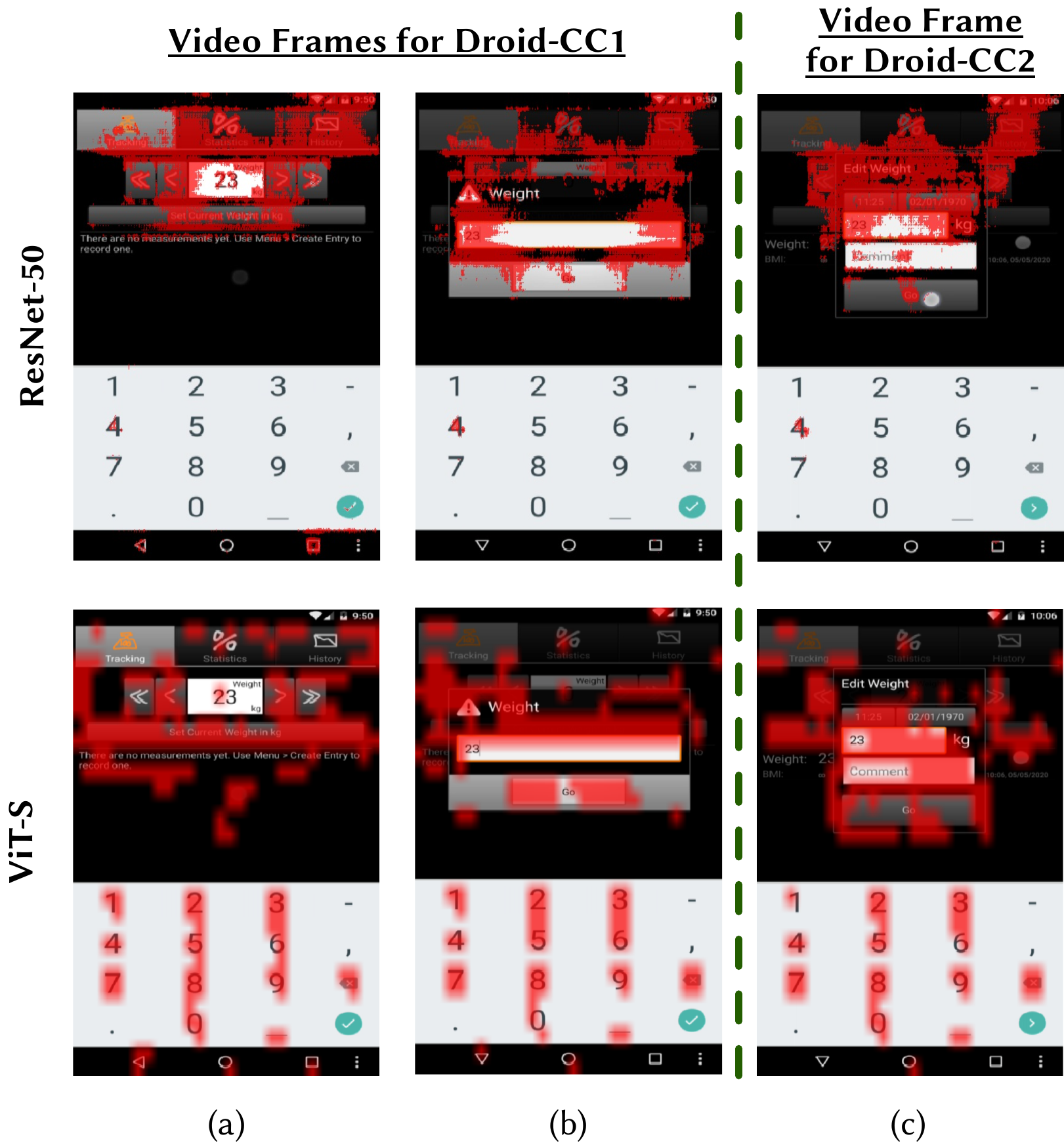} 
\caption{\normalsize{Visualization of ResNet-50 and ViT} on keyframes of video-based bug reports}
\label{fig:visual} 
\end{figure} 

In Fig. \ref{fig:visual}, we show three keyframes of two non-duplicate video-based bug reports from DroidWeight (DROID)~\cite{droidweight}: DROID-CC1 and DROID-CC2. SimCLR fails to distinguish between the videos of these two bugs and mistakenly ranks DROID-CC2 as the first duplicate video of DROID-CC1. The DROID-CC1 video mainly has one trace that generates a new weight record by entering the weight on a pop-up component (Fig. \ref{fig:visual} (b)), while the DROID-CC2 video not only includes the previous trace but also a trace that further edits the recorded weight on another different pop-up component (Fig. \ref{fig:visual} (c)). 
Fig. \ref{fig:visual} illustrates the saliency maps, overlaid over frames from two Droidweight video-based bug reports. We observe that the ViT-S/16 is able to attend to key parts of GUI components that ResNet-50 does not. 
Specifically, for the main screen (a) and \textit{entering weight} screen (b) from videos of DROID-CC1 shown in Fig. \ref{fig:visual}, ResNet-50 and ViT-S/16 are all able to attend well to the objects, but ViT-S/16 pays more attention to the GUI layout information. However, for the \textit{edit weight} screen (c) from videos of DROID-CC2, ResNet-50 has more difficulty in distinguishing between foreground pop-up components and the background. We can see it pays less attention to the lower edges and the bottom part of the foreground component. In contrast, ViT-S/16 effectively attends better to the edges and pays enough attention to the foreground component to help distinguish between (b) and (c), hence improving performance on this specific duplicate detection task.

From this example, there is a clear benefit to the visual nuances learned by ViTs. While here we present one example, after investigating several cases where \janusv outperforms the baseline, we observed this pattern holds, wherein \janusv learned visual representation is able to better capture nuanced visual patterns, such as the difference between two similar pop-ups, or the difference between background and foreground element when menus are displayed.
\looseness=-1

\begin{figure}[t] 
\centering 
\includegraphics[width=0.49\textwidth]{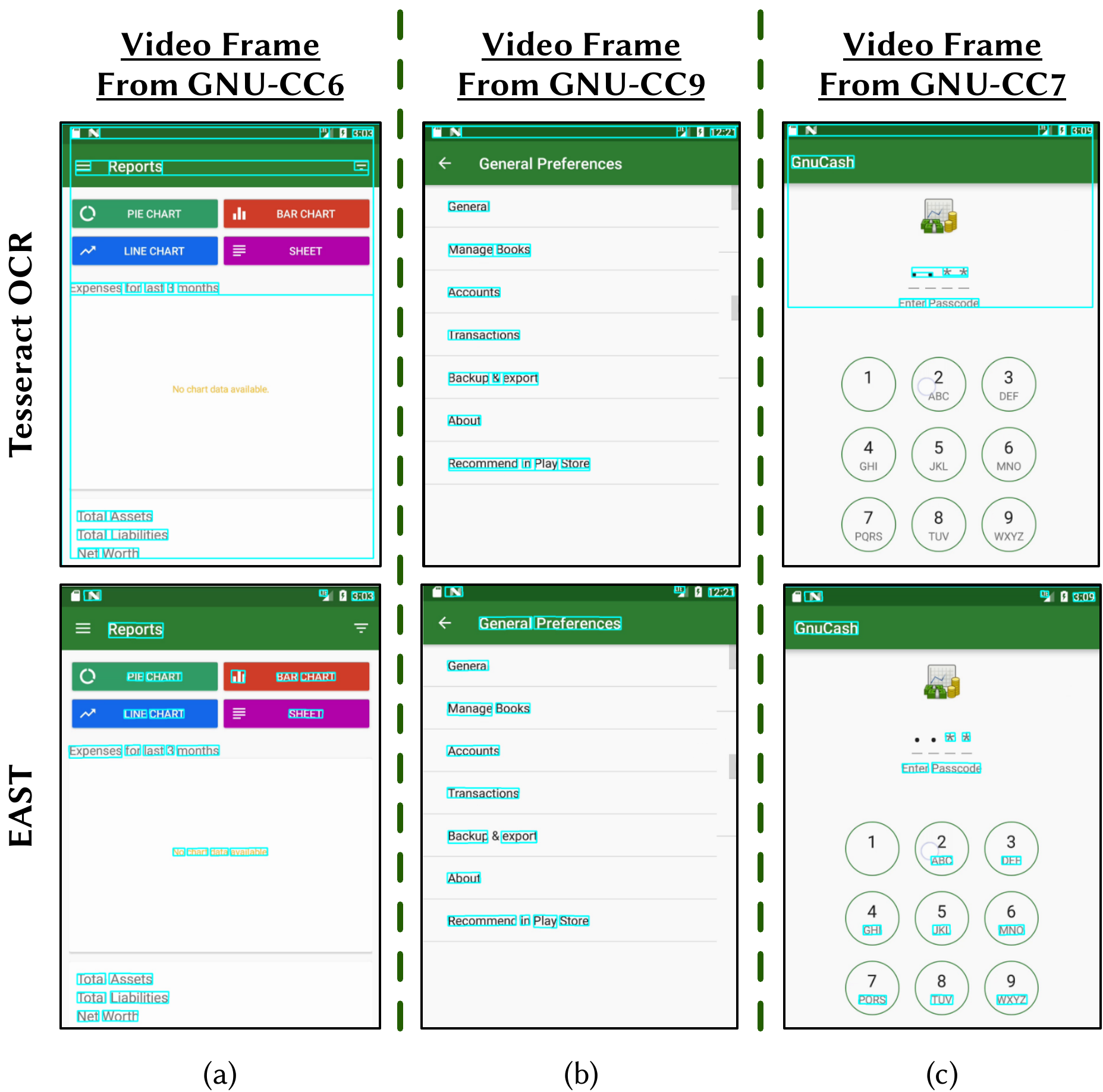}
\caption{\normalsize{Bounding boxes localized by EAST and the Tesseract OCR library on keyframes of video-based bug reports}} 
\label{fig:text} 
\end{figure}

\subsubsection{\textbf{Example 2: Scene-based Text Detection Improves Text Localization}} Textual \tango, which uses  Tesseract OCR is unable to distinguish between similar video reports for a number of bugs, including three bugs from the GNUCash (GNU) app~\cite{gnucash}. Therefore, we visualize the detection bounding boxes of text for three keyframes of these three videos in Fig. \ref{fig:text} for both Tesseract (first row) and EAST~\cite{zhou2017east} (used by \approach). The first report for the GNU-CC6 bug has a main trace that goes to the \texttt{\small balance sheet} screen and checks the sub-account: we show one keyframe for this report in (Fig.~ \ref{fig:text}-(a)), while the second video report for the GNU-CC9 bug navigates to the \texttt{\small General Preferences} screen, as shown in keyframe in (Fig.~ \ref{fig:text}-(b)), and finally, the report for GNU-CC7 changes the password under the \texttt{\small General Preferences} menu, as shown in (Fig.~ \ref{fig:text}-(c)). While these bugs are different, they include many similar screens where keywords are important for differentiation.

As observed in Fig.~\ref{fig:text}, EAST is more accurate than TesseractOCR for GUI component and text detection. 
In Fig.~ \ref{fig:text}-(a), Tesseract OCR fails to localize the text on some buttons (\eg sheet) and the text in brighter colors (\eg Asset). Also, for the keyframe of GNU-CC9 (Fig.~ \ref{fig:text}-(b)), Tesseract misses the text \texttt{\small General Preferences}, making it difficult to distinguish between report GNU-CC9 and GNU-CC7, as they both access various parts of the settings menu. In addition, Tesseract fails to detect the text when it is in low brightness and low contrast regions, including the text on the dialing circles (Fig.~\ref{fig:text}(c)), which also helps with differentiating between GNU-CC9 and GNU-CC7, since GNU-CC7 enters a passcode, but GNU-CC9 only accesses the passcode settings. Thus, the more accurate text extraction of EAST clearly aids in the accurate extraction of key text that can help to differentiate between similar GUI screens.

\section{Threats to Validity}

\subsection{Internal and Construct Validity} 
Beyond the evaluation dataset, the implementation of \aps models and experimental settings represent key validity threats. We controlled as many factors as possible for a fair comparison with the baseline. For instance, we implemented the 4-Codebook approach on both \ap and the baseline, used the same duplicate detection tasks, and measured their performance using well-known metrics in duplicate detection studies.

\subsection{External Validity} 
To improve generalization, we created a new dataset to include $\approx$3k more duplicate detection tasks, for real bugs of different kinds, reported on mobile app issue trackers. These bugs were video recorded by multiple users on various mobile OS versions and did not include touch indicators. We ensured the recorded videos contained different reproduction scenarios for the same bugs. The decisions were made to make our dataset more comprehensive, realistic, and diverse. Our dataset could be improved by considering different app languages or other mobile platforms such as iOS.
\section{Related Work}

\subsection{Duplicate Textual Bug Report Detection}
Many approaches have been proposed to detect duplicate textual bug reports to help developers avoid redundant effort during bug management. Most of the approaches leverage text retrieval techniques to obtain a ranked list of candidate duplicates for a query report~\cite{Runeson2007, Sun2010, Banerjee2013, song2022toward}. Some approaches leverage extra information (fields \cite{Sun2011, Rocha2016, Yang2016}, contexts \cite{Alipour2013, Hindle2015}, execution traces \cite{Wang2008}, \etc) and/or more effective similarity techniques (BM25F~\cite{Sun2011, Tian2012}, topic-modeling \cite{Nguyen2012},  word-embedding \cite{Yang2016}, \etc) to improve the detection. Wang \etal~\cite{Wang2019} proposed SETU, which combines textual bug descriptions with screenshots to detect duplicates, rather than focusing on video reports (as we do).  

\subsection{Automated GUI Understanding for SE}
Various GUI understanding approaches have been proposed to help software engineering tasks related to mobile apps, such as GUI reverse engineering \cite{Beltramelli2018, Chen2018, Moran2020, Zhao2021}, software testing~\cite{BernalCrdenas2020, yu2022universally, Liu2020}, and GUI search \cite{Chen2020_lost, Chen2020_wireframe}. Most of them detect GUI elements first to understand GUI information. Chen \etal~\cite{Chen2020_object} show that deep learning-based object detection models (FasterRCNN \cite{Ren2017}, YOLOv3 \cite{yolov3_arxiv.1804.02767}, and centerNet \cite{Duan2019}) and scene text detector EAST \cite{zhou2017east} outperform old-fashioned detection models~\cite{Nguyen2015} and OCR tool Tesseract~\cite{Smith2007} respectively. However, these models, based on supervised learning, leverage GUI information limited to a few GUI element categories, and the relationships between different elements are not considered, thus lacking an understanding of the entire screen. Fu \etal~\cite{pixel_words/arxiv.2105.11941} therefore attempt to understand the whole screen by considering these relationships, based on the Transformer architecture to detect GUI elements more accurately.

The most closely related work to our own is Cooper \etals~\cite{cooper2021takes}, which proposed the \tango duplicate detector and a dataset to evaluate it. While  \ap leverages the same information from video-based bug reports as \tango does, there are key differences that set \ap apart. First, \ap learns visual features from app GUIs (via distillation and Vision Transformers) which capture GUI layouts more effectively for duplicate detection than \tango, which focuses on learning local GUI features (via contrastive learning and CNNs). Second, \ap learns textual representations of videos that are more useful for duplicate detection, by recognizing and extracting frame text more accurately (via fully neural models rather than heuristic+neural based approaches adopted by \tango). Third, \aps sequential similarity computation, which attempts to align video frames, can be applied to both visual and textual representations, rather than to only visual representations as \tango does. Fourth, the best configuration of \ap combines all three modalities of video information (visual, textual, and sequential), and significantly outperforms the best \tango configuration, on duplicate detection tasks that include both injected and real bugs for a diverse set of mobile apps. Notably, our evaluation dataset is more comprehensive, realistic, and diverse than the one used to evaluate \tango.

\vspace{-0.2cm}
\section{Conclusions}

To assist developers in identifying video-based bug reports that show identical mobile app bugs, we propose \ap, a new approach for duplicate video-based bug report detection. \ap leverages visual, textual, and sequential information from videos via the combination of representation learning, information retrieval, and frame-alignment approaches.  

We evaluated \ap and found that it significantly outperforms an existing duplicate detector. The evaluation considered a new benchmark of 7,290 duplicate detection tasks based on 270 video-based bug reports, drastically extending a prior dataset (with real bugs as opposed to injected bugs from prior work). We conducted ablation experiments and an in-depth qualitative analysis visually showing that \toolname learns a more interpretable hierarchical visual representation and localizes text regions more accurately.

\begin{acks}
This work is supported in part by the following NSF grants: CCF-2311469, CCF-2007246, and CCF-1955853. Any opinions, findings, and conclusions expressed herein are the authors' and do not necessarily reflect those of the sponsors.
\end{acks}

\balance
\bibliographystyle{ACM-Reference-Format}
\bibliography{reference}

\end{document}